\documentclass[aps,a4paper,12pt]{revtex4}
\usepackage{float}
\usepackage{fancyhdr}
\usepackage{color}
\usepackage{graphicx}
\usepackage{epsfig} 
\usepackage{amssymb}
\usepackage{amsmath,amsfonts}
\usepackage{booktabs}
\usepackage{hyperref}
\usepackage{multirow}
\def\bra#1{\mathinner{\langle{#1}|}}
\def\ket#1{\mathinner{|{#1}\rangle}}
\def\normsq#1{\mathinner{\langle{#1}|{#1}\rangle}}
\def\sprod#1#2{\mathinner{\langle{#1}|{#2}\rangle}}

\def\spin#1#2{{\Bigg(\!\begin{array}{c} #1\\[-3mm]#2                   
\end{array}\!\Bigg)}}
\DeclareMathAlphabet{\mathbbmsl}{U}{bbm}{m}{sl}
{\catcode`\|=\active\gdef\Braket#1{\left<\mathcode`\|"8000\let|\bravert {#1}\right>}}
\def\bravert{\egroup\,\vrule\,\bgroup}
\def\Tr{\mathop{\mbox{\normalfont Tr}}\nolimits}

\def\ii{{\rm i}}
\def\ee{{\rm e}}

\def\NC{{\mathbb C}}

\begin{document}

\title{
 The Zeno effect in a quantum computer
}

\author{J. G. Esteve}
 \email{esteve@unizar.es}
 \affiliation{Departamento de F\'{\i}sica Te\'orica, Universidad de Zaragoza,
50009 Zaragoza, Spain}
\affiliation{Instituto de Biocomputaci\'on y F\'{\i}sica de Sistemas
Complejos (BIFI), 50009 Zaragoza, Spain}
  \author{Fernando Falceto}
\email{falceto@unizar.es}
 \affiliation{Departamento de F\'{\i}sica Te\'orica, Universidad de Zaragoza,
50009 Zaragoza, Spain}
\affiliation{Instituto de Biocomputaci\'on y F\'{\i}sica de Sistemas
Complejos (BIFI), 50009 Zaragoza, Spain} 

\begin{abstract} 
 We present a simulation of the quantum Zeno effect (QZE) on a quantum computer
 as an example of the relation between this effect and the bang-bang decoupling 
 method in control theory. Although the true QZE can not be strictly implemented on a 
 quantum computer where all the operations, except the final measurements, must be unitary ones, 
 we can simulate it by coupling the system to a number of ancillas and replacing the projective measurements of the QZE by suitable unitary gates. 
  \end{abstract}

\maketitle

\section{Introduction}\label{sec_intro}

The quantum Zeno ($ Z \eta \nu \omega \nu$) effect (QZE), so named after Misra and Sudarshan \cite{Sudarshan},
is another striking effect of the quantum theory which states that similarly to the Zeno 
paradoxes and the Eleatic philosophy, a quantum system will not evolve from its initial state 
if it is continuously observed (see references \cite{Pascazio1}-\cite{Shimizu} for 
a complete description of the Zeno effect). 
The QZE has been proved  in different physical systems, 
ranging from unstable systems (atomic tunneling and in spontaneous parametric down conversion) 
to Bose-Einstein condensates, 
optical pumping  or in cavity quantum electrodynamics \cite{Shimizu}, and
recently,  there has been also several proposals for using this effect to control decoherence 
or to implement  deterministic logical gates 
\cite{Haroche}.

The key point of the usual QZE is the non-unitarity of the measurement 
operation and the collapse of the wave function after any measurement. 
This characteristic  makes difficult, in principle, to 
reproduce exactly the Zeno effect on a quantum computer, because the quantum gates are unitary and the 
measurement of the qubits in the computer are done, only, at the end of the circuit. 
However, we have learned also that a strong coupling to an external system can play the role of 
a measuring apparatus and in this way, the projective measurements can be substituted by a 
strong interaction with the external system, see  \cite{flp} - \cite{bfgk} 
where it has been proved  
that given a bounded self-adjoint Hamiltonian $H$ and a unitary 
operator $U$ with a spectral representation $U=\sum_n \ee^{i \varphi_n} P_n$ we have:
\begin{equation}
 \lim_{N\to\infty} {\left( U \exp{(\ii \frac{H}{\hbar }\frac{t}{N})}\right)^N}= U^N \exp{(\ii \frac{H_z }{\hbar} t)},
 \label{uno}
\end{equation}
where $H_z$ is the Zeno Hamiltonian defined as the  Hamiltonian projected onto the invariant 
subspaces of $U$:
\begin{equation*}
 H_z=\sum_n P_n H P_n.
\end{equation*}

The equation (\ref{uno}) opens the opportunity to simulate the Zeno effect on a 
quantum computer, using a qubit $\ket{q_0}$ as the system and one or several auxiliary 
qubits as the environment; then applying periodically controlled unitary gates 
between our system and the auxiliary qubits we can simulate the measuring apparatus.
The aim of this paper is to show that a simulation of the QZE 
can be done in a quantum computer using different strategies with different numbers of auxiliary ancillas. In first place we will discuss the general case in the limit of large number of conditioned gates.
Next we will study in more detail two extreme cases. In the
first one we couple the qubit of interest $\ket{q_0}$ to $N$ ancillas (as many as the number of controlled 
operations); then we divide the total evolution of 
the state $\ket{q_0}$ in N steps and at each of these steps we make a controlled 
operation, taking $\ket{q_0}$ as the control qubit  and one ancilla (a different one each time) as the target qubit.
In this way, the final wave function can be written as 
$\ket{\Psi_f}=\ket{0} \otimes  \ket{\varphi_0}  + \ket{1} \otimes \ket{\varphi_1} $ where $\ket{0},\ket{1}$ 
are the states of $\ket{q_0}$ and $ \ket{\varphi_0} ,  \ket{\varphi_1} $ represent the states of the $N$ ancillas. The final 
probabilities $P_s(n)$ of finding the qubit $\ket{q_0}$ in the states $\ket{s}$ 
(for $s=0,1$) are $\normsq{\varphi_s}$ and, in general,
will be the sum of the  probabilities of the different states of the ancillas. 
The second case we shall consider is when we have only one ancilla $\ket{q_1}$. 
Acting as before, all the controlled operations 
will have $\ket{q_0}$  as the control qubit and $\ket{q_1}$ as the target qubit. 
In this case, the final wave function  will be
$\ket{\Psi_f}= \ket{0} \otimes (\alpha_0 \ket{0}+\beta_0 \ket{1})  + 
 \ket{1} \otimes (\alpha_1 \ket{0}+\beta_1 \ket{1}) $  
where the most left qubit is $\ket{q_0}$ and the right qubit is the ancilla $\ket{q_1}$. Then the  probabilities 
are $P_s(N)=|\alpha_s|^2+|\beta_s|^2$. In the following sections we will compute these coefficients when the number of controlled gates goes to infinity and we will show how this compares to the Zeno effect where the number of intermediate measurements is large.

The paper is organised as follows. In Section II we present the notation and conventions for the QZE.
In Section III we obtain the recursion relation for a simulation of the QZE on a qubit using 
several auxiliary ancillas and in Section IV we present the results of a simulation using 
only one ancilla for each qubit of interest. Conclusions and outlooks are discussed in the last Section.

\section{The quantum circuit}\label{sec_system}
\begin{figure}[h]
 \center
  \includegraphics[width=16cm]{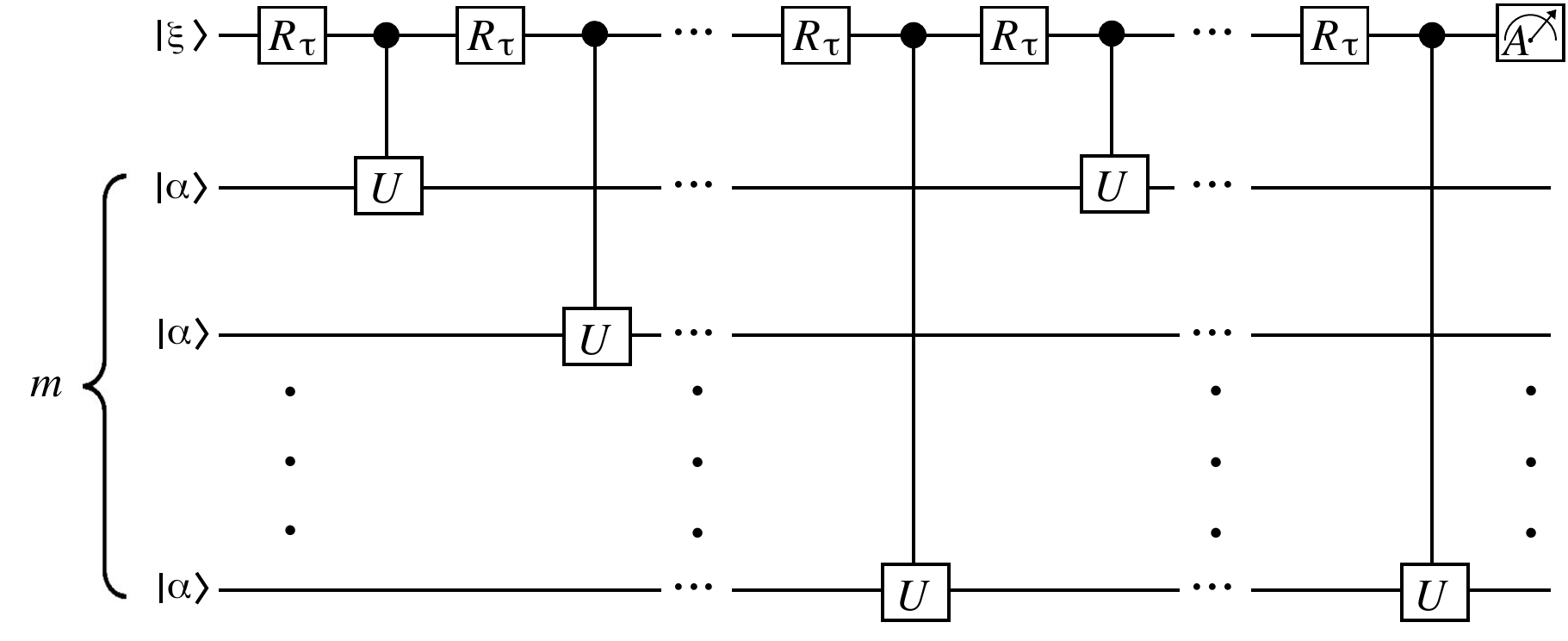}
  \caption{In the figure we represent the circuit for a simulation with $m$ ancillas and $N=mn$ steps. Each step is composed by 
    the  evolution under $R(\tau)$ of the qubit $\ket{q_0}$ and a controlled $C_U(0,b)$ gate controlled by $\ket{q_0}$ and acting each time on a different
   ancilla $\ket{q_b}$. The last operation represents the measurement of the qubit $\ket{q_0}$} 
\label{sistema_u}
 \end{figure}

In this article, we use the computational 
basis $\ket{q_0,q_1,\cdots,q_m}=\ket{q_0}\otimes
\ket{q_1}\otimes\cdots\otimes\ket{q_m}$
where the first qubit $\ket{q_0}$ constitutes the system
and the rest correspond to ancillas.
As it is standard, we denote by $C_{U}(a,b)$ a one qubit unitary operator $U$ acting on $\ket{q_b}$
and controlled by the qubit $\ket{q_a}$,  as for example:
\begin{eqnarray*}
 C_X(0,1) \ket{0 0 }= \ket{0 0 }, \;\;\; C_X(0,1) \ket{0 1 }= \ket{0 1},\\
 C_X(0,1) \ket{1 0 }= \ket{1 1 }, \;\;\; C_X(0,1) \ket{1 1 }= \ket{1 0}.
\end{eqnarray*}
In this section, 
 we shall consider the qubit $\ket{q_0}$, in a initial state $\ket\xi$, coupled with $C_U(0,b)$ gates to $m$ ancillas,
 as it is shown in the picture \ref{sistema_u}.
We denote by $\ket\alpha$ the initial state of the ancillas, assumed all identical.

The qubit $\ket{q_0}$ evolves with Hamiltonian $H$ in $N=nm$ time steps of length $\tau=T/N$ with the unitary operator
$$R(\tau)=\exp{(-\ii \tau H/\hbar)}.$$
Between two consecutive time steps there
is a $C_U(0,b)$ gate that couples $\ket{q_0}$ to one of the ancillas. Once the $m$ ancillas has been coupled to $\ket{q_0}$ the cycle repeats itself $n$ times.

We are interested in measuring $\ket{q_0}$, forgetting the states of the ancillas; i. e. our measurement devices $M$ are of the form
$$M=A\otimes I\otimes\cdots\otimes I ,$$
as it is indicated in the figure. Hence all possible outcomes are encoded in the $2\times2$ final density matrix $\rho_{f}$ which is
obtained by tracing out the factor space corresponding to the $m$ ancillas.

Of course, if $U$ is the identity $\ket{q_0}$ evolves coherently and the final density matrix corresponds to a pure state
$$\rho_{f}=R(T)\ket\xi\bra\xi R(T)^\dagger \, .$$
But we will see that if $U$ is different form $I$ we may have loss of coherence
and some Zeno-like effect in the system. 


In this section we are interested in characterizing its large $N$ properties
for generic Hamiltonian and control gate $C_U(0,i)$. In first place, we shall
consider the case of a finite number of ancillas $m$ and
study the limit when the number of cycles $n$ goes to infinity.

The final density matrix in the computational basis $\{\ket0,\ket1\}$
and for generic $U$
(the meaning of genericity will be specified later) is the following
\begin{equation}\label{rhof}
\rho_{f}=
\begin{pmatrix}
  |\sprod\xi0|^2& B\sprod 0\xi\sprod\xi1\cr B^*\sprod 1\xi\sprod\xi0
  &|\sprod\xi1|^2
\end{pmatrix}
+O(1/N) ,
\end{equation}
with
\begin{equation}\label{rhof1}
B=\bra\alpha U^n\ket\alpha^m
\exp{[-\ii T (\bra 0 H\ket 0 - \bra 1 H\ket 1)/\hbar]}.
\end{equation}
The two factors that compose $B$ have the following meaning: the second one gives account of the evolution with the Zeno Hamiltonian in the computational basis, which is invariant under $C_U$
$$H_z=P_0HP_0+P_1HP_1 ,$$
while the first factor measures the loss of coherence of the final state. In fact if one computes the purity of $\rho_f$ one gets
$$\Tr(\rho_f^2)=1-2|\!\sprod0\xi\!\sprod1\xi\!|^2
(1-|\!\bra\alpha U^n\ket\alpha\!|^{2m})+ O(1/N)$$
and we have a pure state if $\bra\alpha U^n\ket\alpha$ is a phase, which occurs  when the ancilla is initially an eigenvector of $U^n$.
If this is not the case and $|\!\bra\alpha U^n\ket\alpha\!|<1$, the purity decreases exponentially with the number of ancillas $m$ and it approaches the minimum value $|\!\sprod0\xi\!|^4+|\!\sprod1\xi\!|^4$ when the number of ancillas goes to infinity.

As we already mentioned, (\ref{rhof}) holds for generic $U$ which in our case means
$$\det(U^{\otimes m}-I)\not=0.$$
That is, the  $m$-th tensor power of $U$, $U\otimes\cdots\otimes U$, should not have eigenvalue one. For instance, the multiplication by an irrational (in $2 \pi$ units) phase is always allowed, but a special unitary operator is not permitted if $m$ is even.

In order to prove the result we shall expand the state of the ancillas in a basis of eigenvectors of $U$, $\{\ket{\varphi_+},\ket{\varphi_-}\}$ with eigenvalues $\ee^{\ii\varphi_+}
,\ee^{\ii\varphi_-}$ respectively. For the q-bit $\ket{q_0}$, however, it is convenient to use the computational basis. Their expressions in the respective basis are 
$$\ket\alpha= \beta_+\ket{\varphi_+} +\beta_-\ket{\varphi_-},\qquad\ket\xi=x\ket0+y\ket1.$$

Now we construct the quantum circuit of fig \ref{sistema_u} step by step.
Consider the first ancilla in any of the basic states  $\ket{\varphi_1}\in\{\ket{\varphi_+},
\ket{\varphi_-}\}$ then after the first step ($R_\tau$ and $C_U$)
the initial state $\ket\xi\ket{\varphi_1}$
transforms into $\ket{\xi'}\ket{\varphi_1}$
with
$$\ket{\xi'}=(\ket0,\ket 1)M^{(1)}\begin{pmatrix}x\cr y\end{pmatrix}
  \qquad\mbox{and}
  \quad M^{(1)}=
  \begin{pmatrix}
    \bra0 R_\tau\ket0&&
\bra0 R_\tau\ket1\cr
\ee^{\ii\varphi_1}
\bra1 R_\tau\ket0&&
\ee^{\ii\varphi_1}
\bra1 R_\tau\ket1
\end{pmatrix} .$$
  If we expand $M^{(1)}$ in powers of $\tau=\frac T{N}$ we obtain
  $$M^{(1)}=
  \begin{pmatrix}
    1-\ii \tau a^{(1)}&  \tau b^{(1)}\cr
    \tau c^{(1)}& \ee^{\ii\varphi_1}(1-\ii \tau d^{(1)})
  \end{pmatrix}+O(\tau^2) ,
  $$
  where
  $$a^{(1)}=\frac{1}\hbar
  \bra0 H\ket0,\   d^{(1)}=\frac {1}\hbar\bra1 H\ket1.$$
  
  Therefore, after completing the first $m$ steps, the initial state $\ket\xi\ket{\varphi_1}\cdots\ket{\varphi_m}$ transforms into $\ket{\xi^{(m)}}\ket{\varphi_1}\cdots\ket{\varphi_m}$ where
$$\ket{\xi^{m)}}=(\ket0,\ket 1)M^{(m)}\begin{pmatrix}x\cr y\end{pmatrix} ,$$
    with
    
$$M^{(m)}=
  \begin{pmatrix}
    (1-\ii\tau a^{(1)})^m&  \tau b^{(m)}\cr
    \tau c^{(m)}& \ee^{\ii(\varphi_1+\cdots+\varphi_m)}(1-\ii\tau d^{(1)})^m
  \end{pmatrix}+O(\tau^2),
$$
and, as we shall see,  all we need to know about $b^{(m)}$ and $c^{(m)}$ is that they
  are finite in the large $n$ (or little $\tau=T/(mn)$\,) limit.

  Now, it is clear that in order to complete the quantum circuit of Figure \ref{sistema_u}
  we must take the $n$-th power of $M^{(m)}$. But it is easy to see that if $\ee^{\ii(\varphi_1+\cdots+\varphi_m)}\not=1$
  $$ M^{(nm)}= (M^{(m)})^n =
  \begin{pmatrix}  \ee^{-\ii T a^{(1)}} & 0\cr
    0&  \ee^{\ii n(\varphi_1+\cdots+\varphi_m)-\ii T d^{(1)}}
  \end{pmatrix}+O(1/n).
  $$
  This is because, assuming  $\ee^{\ii(\varphi_1+\cdots+\varphi_m)}\not=1$, the eigenvalues of $M^{(m)}$ are, up to terms of second order in $\tau$, its diagonal entries and the change of basis matrix is the identity up to terms of order $\tau$. Therefore, to compute $M^{(nm)}$ at zero order we must rise the diagonal entries to the $n$-th power (which produces the finite contribution specified above) and we take the identity for the change of basis.

 From the previous result, summing to all possible choices of eigenvectors $\ket{\varphi_i}\in\{\ket{\varphi_+}, \ket{\varphi_-}\}$ for any ancilla with the corresponding coefficients
 $\beta_i\in \{\beta_+, \beta_-\}$ and tracing out the degrees of freedom corresponding to the ancillas, we can get the final reduced density
 y matrix
\begin{eqnarray}
\rho_f&=&\sum_{\{(\beta_i,\varphi_i)\}}|\beta_1|^2\cdots|\beta_m|^2\\
&&\begin{pmatrix}  \ee^{-\ii T a^{(1)}} & 0\cr
    0&  \ee^{\ii n(\varphi_1+\cdots+\varphi_m)-\ii T d^{(1)}}\end{pmatrix}\begin{pmatrix}|x|^2&xy^*\cr yx^*&|y|^2\end{pmatrix}
    \begin{pmatrix}  \ee^{-\ii T a^{(1)}} & 0\cr
    0&  \ee^{\ii n(\varphi_1+\cdots+\varphi_m)-\ii T d^{(1)}}\end{pmatrix}^\dagger\!+O(\frac1n),\nonumber
 \end{eqnarray}
 where the sum extends to pairs 
 $(\beta_i,\varphi_i)\in\{(\beta_+,\varphi_+),(\beta_-,\varphi_-)\}$ for $i=1,\dots,m$. 
 
 If, finally, we take into account the definitions of $a^{(1)}$ and $d^{(1)}$, the relations
 $x=\sprod0\xi$, $y=\sprod1\xi$ and  the identities
 $$\sum_{\{(\beta_i,\varphi_i)\}}|\beta_1|^2\cdots|\beta_m|^2=1,\quad \sum_{\{(\beta_i,\varphi_i)\}}|\beta_1|^2\cdots|\beta_m|^2
\ee^{\ii n(\varphi_1+\cdots+\varphi_m)}=(\bra\alpha U^n\ket\alpha)^m,$$
we obtain the result presented in (\ref{rhof}).


We will discuss now the opposite case of taking $m$ to infinite, actually we shall consider $n=1$ and, therefore $N=m$. 
The quantum circuit is composed of $N$ ancillas and everyone is acted only once. In this case the previous proof is 
not valid any more and we must change our strategy. 

We write the initial state as
\begin{equation}
  \ket{\Psi_0} = \ket{\xi}\otimes\ket\alpha\otimes \cdots \otimes \ket\alpha
  \equiv \ket{\xi}\otimes \ket\alpha^N .
\end{equation}
After $k$ steps, the wave function $\ket{\Psi_k}=\prod_{j=1}^k (C_
U(q_0,q_j) R_\tau)\ket{\Psi_0}$ 
is
\begin{equation}
   \ket{\Psi_k} = 
   (\ket{0}\otimes\ket{\phi_k}
   +\ket{1}\otimes\ket{\eta_k})\otimes\ket\alpha^{N-k}, 
\end{equation}
where $\ket{\phi_k}, \ket{\eta_k}\in(\NC^{2})^{\otimes k}$ are wave functions that represent the 
states of the first $k$ ancillas. They are not orthogonal, in general,  but satisfy $\normsq{\phi_k}+\normsq{\eta_k}=1$. 
If we go one step further, the new wave function is
\begin{equation}
  \ket{\Psi_{k+1}} = C_U(q_0,q_{k+1}) R_\tau\ket{\Psi_k}=
  (\ket{0}\otimes \ket{\phi_{k+1}}
  +\ket{1}\otimes\ket{\eta_{k+1}}
  )\otimes  \ket{\alpha}^{N-k-1} 
,
\end{equation}
with
\begin{eqnarray*}
 \ket{\phi_{k+1}}&=&
 (\Lambda_{00} \ket{\phi_k} +\Lambda_{10}\ket{\eta_k})\otimes\ket{\alpha} ,\\
 \ket{\eta_{k+1}}&=& (\Lambda_{01} \ket{\phi_k} + \Lambda_{11} \ket{\eta_k})\otimes\ket{\alpha'},
\end{eqnarray*}
where $\ket{\alpha'}=U\ket\alpha$ and $\Lambda_{ss'}=\bra sR_\tau\ket{s'}$, with $s,s'=0,1$.

Therefore the following recursion relations hold:
\begin{eqnarray}
  \normsq{\phi_{k+1}}&=&
  |\Lambda_{00}|^2\normsq{\phi_{k}}+
  |\Lambda_{10}|^2\normsq{\eta_{k}}+
  \Lambda_{00}^*\Lambda_{10}\sprod{\phi_k}{\eta_k} +
  \Lambda_{10}^* \Lambda_{00}\sprod{\eta_k}{\phi_k},
\label{itera1}\\
\normsq{\eta_{k+1}}&=&
|\Lambda_{01}|^2\normsq{\phi_{k}}+
|\Lambda_{11}|^2\normsq{\eta_{k}}+
\Lambda_{01}^*\Lambda_{11}\sprod{\phi_k}{\eta_k} +
\Lambda_{11}^*\Lambda_{01}\sprod{\eta_k}{\phi_k},
\nonumber\\
 \sprod{\phi_{k+1}}{\eta_{k+1}}&=&
 \sprod{\alpha}{\alpha'} 
 \big( 
 \Lambda_{00}^*\Lambda_{01}\normsq{\phi_{k}}+\Lambda_{10}^*\Lambda_{11}\normsq{\eta_{k}} + \Lambda_{00}^*\Lambda_{11}\sprod{\phi_k}{\eta_k}+
 \Lambda_{10}^*\Lambda_{01}\sprod{\eta_k}{\phi_k}\big),\nonumber
 \\
 \sprod{\eta_{k+1}}{\phi_{k+1}}&=&
 \sprod{\alpha'}{\alpha} 
 \big( 
 \Lambda_{01}^*\Lambda_{00}\normsq{\phi_{k}}+\Lambda_{11}^*\Lambda_{10}\normsq{\eta_{k}} + \Lambda_{01}^*\Lambda_{10}\sprod{\phi_k}{\eta_k}+
 \Lambda_{11}^*\Lambda_{00}\sprod{\eta_k}{\phi_k}\big).\nonumber
 \end{eqnarray}
 Or in a more compact notation
 \begin{equation}\label{recurrence}
 \begin{pmatrix}
 \normsq{\phi_{k+1}}\\
 \normsq{\eta_{k+1}}\\
 \sprod{\phi_{k+1}}{\eta_{k+1}}\\
 \sprod{\eta_{k+1}}{\phi_{k+1}}
 \end{pmatrix}
 =
{ \cal M}
 \begin{pmatrix}
 \normsq{\phi_{k}}\\
 \normsq{\eta_{k}}\\
 \sprod{\phi_{k}}{\eta_{k}}\\
 \sprod{\eta_{k}}{\phi_{k}}
 \end{pmatrix} ,
 \end{equation}
 with
 \begin{equation}\label{recurr_matrix}
{\cal M}
 =
 \begin{pmatrix}
   |\Lambda_{00}|^2 &
   |\Lambda_{10}|^2&
   \Lambda_{00}^*\Lambda_{10}&
   \Lambda_{10}^*\Lambda_{00}
 \\
 |\Lambda_{01}|^2&
 |\Lambda_{11}|^2&
 \Lambda_{01}^*\Lambda_{11}&
 \Lambda_{11}^*\Lambda_{01}
\\
z\Lambda_{00}^*\Lambda_{01}& 
z\Lambda_{10}^*\Lambda_{11}&
z\Lambda_{00}^*\Lambda_{11}&
z\Lambda_{10}^*\Lambda_{01}
 \\
 z^*\Lambda_{01}^*\Lambda_{00}&
 z^*\Lambda_{11}^*\Lambda_{10}&
 z^*\Lambda_{01}^*\Lambda_{10}&
 z^*\Lambda_{11}^*\Lambda_{00}
 \end{pmatrix} ,
 \end{equation}
 and $z=\sprod{\alpha}{\alpha'}=\bra\alpha U\ket\alpha$.
 
The final density matrix in the computational basis, after $N=m$ steps (recall that 
we take $n=1$) and tracing out the ancillas degrees of freedom, is

$$\rho_f=
\begin{pmatrix}
 \normsq{\phi_{N}}&\sprod{\phi_{N}}{\eta_{N}}\\
 \sprod{\eta_{N}}{\phi_{N}}&\normsq{\eta_{N}} 
\end{pmatrix} ,
$$
whose entries can be obtained if we solve the iteration with the initial state
for the qubit $\ket{q_0}$ given by $\ket\xi= x\ket0+y\ket1$. That is
$$
\begin{pmatrix}
 \normsq{\phi_{N}}\\
 \normsq{\eta_{N}}\\
 \sprod{\phi_{N}}{\eta_{N}}\\
 \sprod{\eta_{N}}{\phi_{N}}
 \end{pmatrix}
={\cal M}^N
\begin{pmatrix}
|x|^2\\
|y|^2\\
yx^*\\
xy^*
 \end{pmatrix} ,
$$

We are interested in the large $N$ limit of $\rho_f$. For that it is enough to 
consider the expansion of $\cal M$ to first order in $\tau=T/N$ i. e.
$$
{\cal M}
 =
 \begin{pmatrix}
 1 & 0&\tau c^{(1)}&\tau {c^{(1)}}^*
 \\
 0&1&\tau {b^{(1)}}^*
 & \tau {b^{(1)}}
\\
\tau z {b^{(1)}}& 
\tau z {c^{(1)}}^* &
z(1+\ii\tau (a^{(1)}-d^{(1)}))&
0
 \\
 \tau z^*{b^{(1)}}^*&
 \tau z^*{c^{(1)}}&
 0&
 z^*(1-\ii\tau (a^{(1)}-d^{(1)}))
 \end{pmatrix}
 +O(\tau^2) .
 $$

 Now, we are in a situation similar to the previous case. For $z\not=1$ the 
 eigenvalues of $\cal{M}$ up to $O(\tau^2)$ terms are its diagonal entries 
 while the extra diagonal elements of $\cal M$ contribute only, at order 
 $O(\tau)$, to the change of basis matrix. Then, considering all toghether we can write
 $${\cal M}^N=
 \begin{pmatrix}
 1 & 0&0&0
 \\
 0&1&0&0
\\
0&0&z^N\ee^{\ii T(a^{(1)}-d^{(1)})}&0
 \\
 0&0&0&z^{*^N}\ee^{-\ii T(a^{(1)}-d^{(1)})}
  \end{pmatrix}
 +O\left(1/N\right) .
 $$

 Summarizing the final density matrix for $\ket{q_0}$ can be written
\begin{equation*}
\rho_{f}=
\begin{pmatrix}
  |\sprod\xi0|^2& B\sprod 0\xi\sprod\xi1\cr B^*\sprod 1\xi\sprod\xi0
  &|\sprod\xi1|^2
\end{pmatrix}
+O(1/N) ,
\end{equation*}
with
$$
B=\bra\alpha U\ket\alpha^N
\exp{[-\ii T (\bra 0 H\ket 0 - \bra 1 H\ket 1)/\hbar]}.
$$
Observe that this is exactly the same expression that we obtained in (\ref{rhof1}) 
particularized to the case $n=1$. However, the conditions under which this relation 
holds are more general. It is enough to require
$\det(U-I)\not=0$
to guarantee that the proof is valid for every initial state of the ancilla 
(note that $\det(U^{\otimes m}-I)\not=0$ implies $\det(U-I)\not=0$).

One is tempted to conjecture that (\ref{rhof}) is valid in the large $N$ limit whether  
$n$, $m$ or both tend to infinity, but this is something that at present we can not prove.
 
Coming back to the particular case of $n=1$ we observe that if the initial state of 
the ancilla is not an eigenstate of $U$, $|\bra{\alpha} U\ket\alpha|<1$ and $B$ vanishes when $N\to\infty$. Therefore
$$\lim_{N\to\infty}
\rho_f=
\begin{pmatrix}
  |\sprod\xi0|^2& 0 \cr 0
  &|\sprod\xi1|^2
\end{pmatrix}
+O(1/N) ,
$$
and, on the one hand side,  the final state loses all coherence, it is merely an
statistical mixture of $\ket0$ and $\ket1$ states. On the other hand, the evolution 
given by $H$ is completely freezed. 

This is reminiscent of the Zeno effect only that now we make a single measurement and 
the final states is not determined by the measurement apparatus $A$ but by the control states of the interaction with the ancillas.

In order to compare with the Zeno effect it is convenient to refine this somehow rude 
zero order approximation and compute terms linear in $1/N$ or higher orders. This is 
accomplished for some particular cases in the following sections.


\section{Simulation with multiple auxiliary qubits}
As a first example we shall study the proposal \cite{Panigrahi} to slowing down the evolution 
of the qubit $\ket{q_0}$ using CNOT  gates and several auxiliary qubits. We will consider 
the case $n=1$, $m=N$ of the previous section, i. e.  apart of the qubit of 
interest $\ket{q_0}$ our system has other $N$ ancillas ($m=N$), initially in the state $\ket0$,
on which the controlled gates act only once ($n=1$).
Without lose of generality, the 
Hamiltonian that describes the Rabi oscillation of the qubit can be taken
$H(\theta)=\hslash\omega (c_\theta Y+ s_\theta Z)$, where $c_\theta$ and 
$s_\theta$ stand respectively for the cosine and the sine of the angle $\theta$ of the rotation axis with $Y$. 

As in the previous section we divide the evolution time $T$ into $N$ equal steps of duration $\tau=T/N$.
Therefore the evolution operator between two consecutive controlled gates is
$$R_\tau(\theta)=\exp{(-\ii H \tau/\hslash)}=
\left( \begin{array}{c c}
        \cos(\omega \tau)-\ii s_\theta\sin(\omega\tau)&-c_\theta\sin(\omega\tau)\\
        c_\theta\sin(\omega\tau)&\cos(\omega t)+\ii s_\theta\sin(\omega\tau)
       \end{array}
\right).
$$
which corresponds to a rotation of angle $\omega \tau$ in the axis $c_\theta Y+s_\theta Z$.

Notice that according to the criteria of the previous section $U=X$ is not a generic operator as $\det(X-I)=0$, therefore the results obtained there do not necessarily apply \footnote{In fact, as we will see, the only requirement for our procedure to work is that the initial state of the ancilla $\ket{\alpha}$ and $U\ket{\alpha}$ are orthogonal. Therefore by replacing $X$ with $\ee^{\ii\varphi}X$ we have similar results with a generic operator.}.  The interesting aspect of this kind of quantum circuits, apart from the proposal in \cite{Panigrahi} mentioned above, is that they are exactly solvable, in terms of elementary functions, for any $N$. 

\begin{figure}[H]
 \center
  \includegraphics[width=10cm]{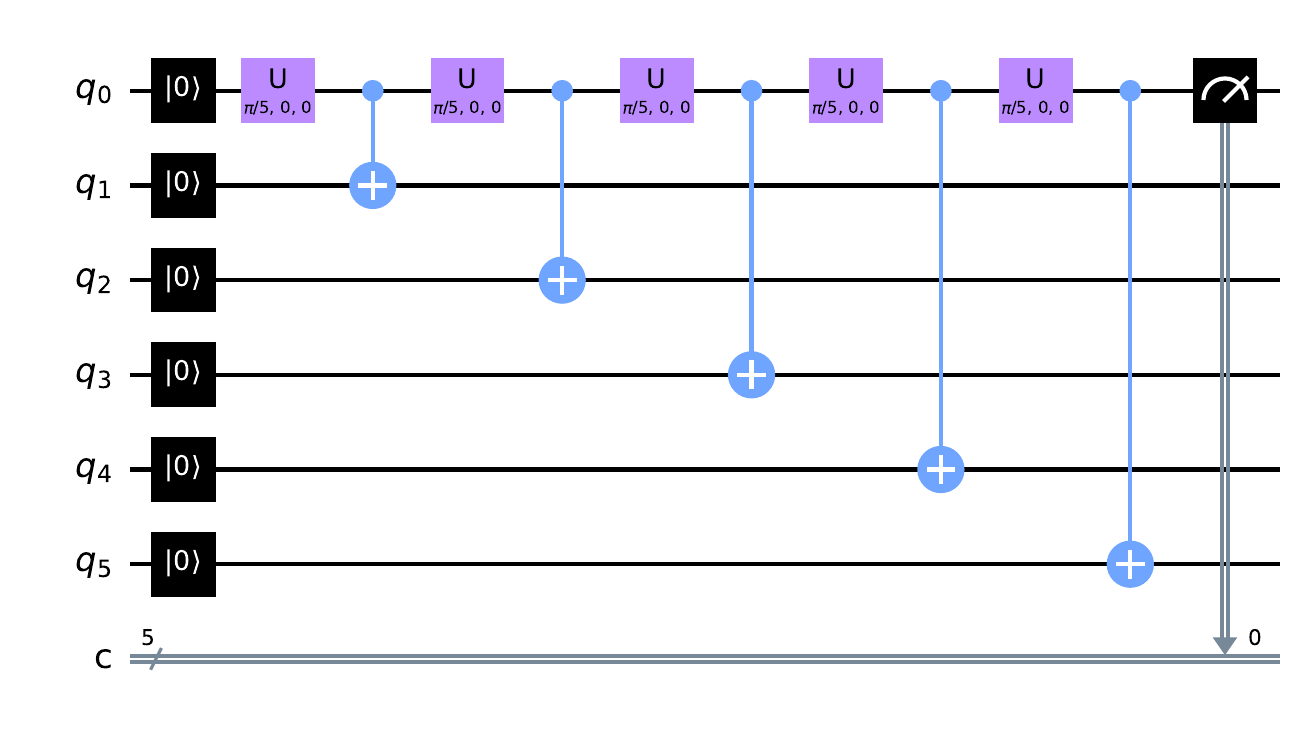}
  \caption{In the figure we represent the circuit for a simulation with $5$ ancillas and $5$ steps. Each step is composed by 
   the  evolution of the qubit $\ket{q_0}$ and a $C_X(0,i)$ gate controlled by $\ket{q_0}$ and acting each time on a different
   ancilla $\ket{q_i}$.  
Finally we measure the qubit $\ket{q_0}$.
   In this example we take $\ket\xi=\ket{\alpha}=\ket0$ and the evolution operator corresponds to $\theta=0$.} 
\label{Circ_x}
 \end{figure}

As we mentioned before, we take initially  all the ancillas in the state $\ket0$ while the q-bit of interest is taken in a general state $\ket\xi=x\ket0+y\ket1$. The initial state is therefore
$$\ket{\Psi^\xi_0}=\ket\xi\bigotimes_{n=1}^N\ket0.$$
Before solving for this general state we shall consider the particular case in which the q-bit of interest is in the computational basis, i.e. 
$\ket\xi=\ket{q_0}$ with $q_0=0,1$. Then it is easy to check that after the first step ($R_\tau$ and CNOT gates) we obtain
$$\ket{\Psi_1^{q_{_0}}}=\sum_{q_{_1}\in\{0,1\}}\ket{q_1}\otimes f(q_0,q_1)\ket{q_1}\bigotimes_{n=2}^{N}\ket{0},$$
where $f$ stands for the matrix elements of $R_\tau$ (or rather its transpose), that is
$$f(0,0)=f(1,1)^*=\cos(\omega \tau)-\ii s_\theta\sin(\omega\tau),\qquad
f(0,1)=-f(1,0)=c_\theta\sin(\omega\tau).$$

Likewise, after the second step we get
$$\ket{\Psi^{q_{_0}}_2}=\sum_{q_{_1}\!,\,q_{_2}\in\{0,1\}}\ket{q_2}
\otimes f(q_0,q_1)\ket{q_1}\otimes f(q_1,q_2)\ket{q_2}
\bigotimes_{n=3}^{N}\ket0.$$

The iteration is therefore immediate and the final state of the full quantum circuit is
$$\ket{\Psi^{q_{_0}}_N}=\sum_{\vec{\, q}\,\in\{0,1\}^N}\ket{q_N}
\bigotimes_{n=1}^{N}f(q_{n-1},q_n)\ket{q_n},\quad q_0=0,1$$

For the general initial state $\ket\xi$ we can apply the superposition principle to get
$$\ket{\Psi^{\xi}_N}= x \ket{\Psi^{0}_N}
+  y \ket{\Psi^{1}_N},$$
and the density matrix, after tracing out the ancillas, reads
$$\rho(T)=\sum_{\vec{\, q}\,\in\{0,1\}^N}
|xf(0,q_1)+yf(1,q_1)|^2 \prod_{n=2}^N |f(q_{n-1},q_n)|^2 \ket {q_N}\bra {q_N}.
$$
The first observation is that the density matrix is diagonal in the computational basis, it represents a classical statistical mixture and quantum coherence is completely lost.  If we are interested in the first entry $P_0(N,\tau)=\bra0\rho(T)\ket0$
(of course, the other non vanishing entry is equal, exchanging $x$ and $y$) we
must take $q_N=0$ and we can decompose the sum so that
\begin{equation}\label{rho00}
P_0(N,\tau)=\sum_{q_{_1}=0,1}|xf(0,q_1)+yf(1,q_1)|^2 \Theta(q_1),
\end{equation}
with
$$\Theta(q_1)=\sum_{(q_{_2}, \dots,\, q_{_{N-1}})\in\{0,1\}^{N-2}}
\ \prod_{n=2}^{N-1} \left |
f(q_{n-1},q_n)\right|^2\left|f(q_{N-1},0)\right|^2.$$
Now we can compute the product by observing that $|f(0,0)|^2=|f(1,1)|^2$ and $|f(0,1)|^2=|f(1,0)|^2$ 
and this implies that if $q_1=0$ we have an even number of factors of the form $|f(1,0)|^2$ while we have an odd number for $q_1=1$. Explicitly
$$\Theta(q_1)=
\begin{cases}
  \displaystyle
  \sum_{\scriptsize\begin{matrix}
      \\[-2.7em]k=0\\[-1em] k\ \mbox{even}
  \end{matrix}}^{N-1}
  {\spin{N-1}{k}}\ 
  |f(0,0)|^{2(N-1-k)}|f(1,0)|^{2k},&q_1=0
      \cr
   \displaystyle\sum_{\scriptsize\begin{matrix}
      \\[-2.7em]k=0\\[-1em] k\ \mbox{odd}
  \end{matrix}}^{N-1}
  {\spin{N-1}{k}}\ 
  |f(0,0)|^{2(N-1-k)}|f(1,0)|^{2k},&q_1=1.
      \end{cases}
  $$    
  And using the definition of $f$ we can evaluate the sum to give
  $$\Theta(q_1)=\frac12 \left[1+(-1)^{q_{_1}}\left(1-2c_\theta^2\sin^2(\omega\tau)\right)^{N-1}\right].$$
  Now, inserting this expression into (\ref{rho00}) we can obtain a closed form for 
  the probability of finding, after $N$ steps or at time $T=N\tau$, the final state at $\ket0$. 
 \begin{eqnarray*}
 P_0(N,\tau)&=&\frac12\left[1+(|x|^2-|y|^2)\left(1-2c_\theta^2\sin^2(\omega\tau)\right)^N\right]\cr
 &&\hskip -1cm-
 \Big[(x^*y+xy^*)\cos(\omega\tau)+\ii(x^*y-xy^*)s_\theta\sin(\omega\tau)\Big]
 \left(1-2c_\theta^2\sin^2(\omega\tau)\right)^{N-1}c_\theta\sin(\omega\tau).
 \end{eqnarray*}

 In order to compare with the Zeno effect, where we consider successive projective measurement on $\ket0$,
 we must take the initial state $\ket\xi=\ket0$ also. Then the previous expression reduces to
 \begin{eqnarray}\label{P_0T}
   P_0(N,\tau)&=&\frac12\left[1+\left(1-2c_\theta^2\sin^2(\omega\tau)\right)^N\right].
 \end{eqnarray}

In the case of a true Zeno effect, we observe the qubit $\ket{q_0}$ every instant $t=n \tau$ with $n=1,2\dots,N$ and $\tau=T/N$. 
For the first interval $\tau$, 
if the initial state is $\ket{0}$ then
$\ket{\Psi(\tau)}=\left(\cos(\omega\tau)-
\ii s_\theta\sin(\omega\tau)\right)\ket{0}+
c_\theta\sin(\omega\tau)\ket{1}$
and since after the observation the wave function collapses 
to the measured state, the probability $Z_0(n,\tau)$ that the qubit has remained always in the state $\ket{0}$ 
after $n$ observations is
$Z_0(n,\tau)=\left|\cos(\omega\tau)-
\ii s_\theta\sin(\omega\tau)\right|^{2 n}=\left(1-c_\theta^2\sin^2(\omega\tau)\right)^n$ 
after $n=N$ steps and in the limit $N>>1$ we have
$Z_0(N,\tau)\approx \exp{(-c^2_\theta\omega^2T^2/N)}$ up to terms $O(1/N^3)$ 
(as an example, for $c_\theta=1, \omega T=\pi/2$ the probability of measuring 
$\ket0$ without intermediate observations is $0$, while $Z_0(10,\tau)=0.78$  or 
$Z_0(100,\tau)=0.975$  showing clearly the effect of the observations). 
Now, comparing $P_0(N,\tau)$  with the probability of the true Zeno effect $Z_0(N,\tau)$, 
where $T=N\tau$ for $N>>1$, we obtain: 
\begin{eqnarray}
 Z_0(N,T/N)=1-\frac{(c_\theta\omega T)^2}{N}+ \frac{(c_\theta\omega T)^4}{2 N^2}+ O(1/N^3) ,\\
 P_0(N,T/N)=1-\frac{(c_\theta\omega T)^2}{N}+ \frac{(c_\theta\omega T)^4}{ N^2}+ O(1/N^3),
\end{eqnarray}
and we can see that the difference is of order $1/N^2$, consequently for high values of $N$, the circuit of Figure \ref{Circ_x} 
can be considered a  good approximation to the Zeno effect. It should be noted that 
$P_0(N,\tau) > Z_0(N,\tau)$ for all $N$. We shall show below the reason for this inequality.

\begin{figure}[h]
 \center
  \includegraphics[width=9cm]{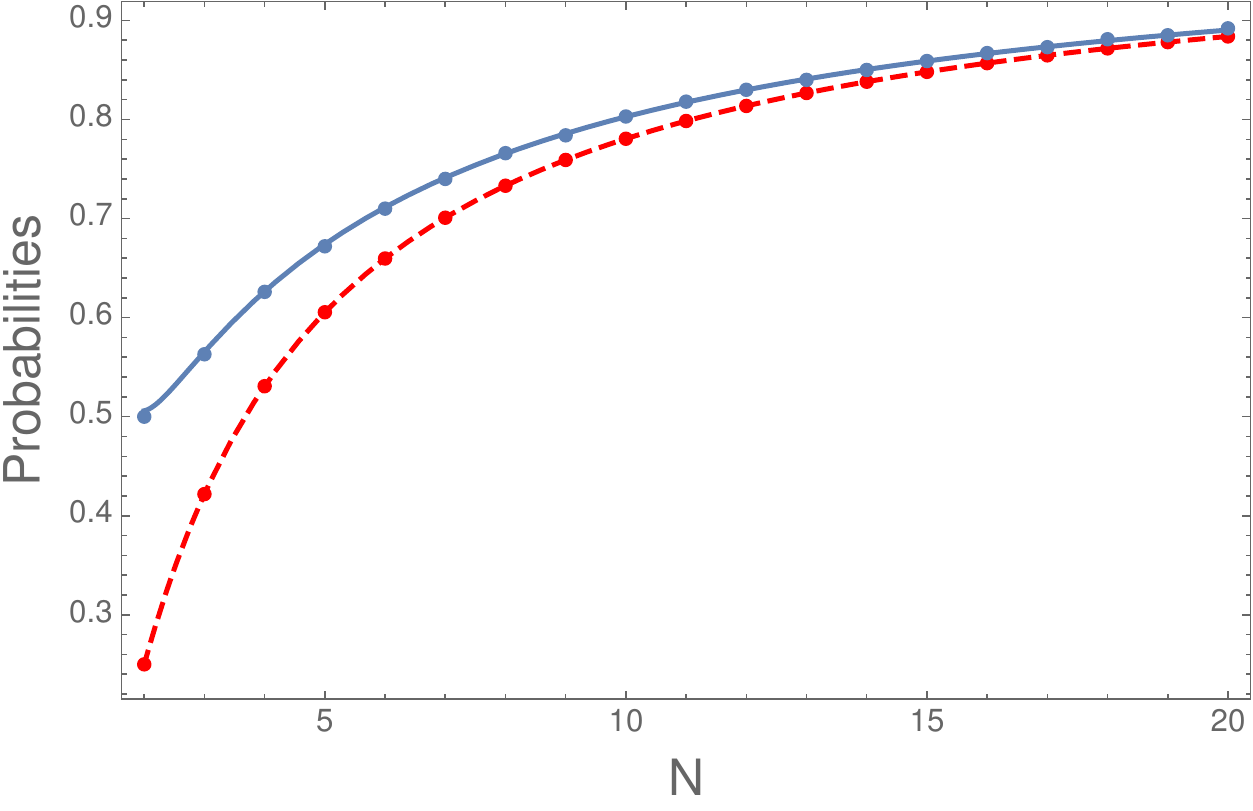}
  \caption{Probabilities $P_0(N,\tau)$ (points joined by a continuous line) and $Z_0(N,\tau)$ 
  (points joined by a dashed line) of finding 
  the qubit $\ket{q_0}$ in the state $\ket{0}$ as a function of the number of ``measurements'' 
  $N$ for $\omega T=\pi/2$, i. e. $\tau=\pi/(2N\omega)$. The initial state is $\ket{\xi}=\ket{0}$  and $\theta=0$. } 
\label{ComparaZ_CX}
 \end{figure}
In the Figure \ref{ComparaZ_CX}  we can see the probabilities $P_0(N,\tau)$ 
(points joined by a continuous line) and $Z_0(N,\tau)$  (points 
joined by a dashed line) for  $\omega T=\pi/2$, which implies $\tau=T/N=\pi/(2N\omega)$, $\theta=0$ 
and several values of $N\in [2,20]$. The points for $Z_0(N,\tau)$ are obtained from 
the equation $Z_0(N,\tau)=\cos^{2N}(\omega\tau)$ and the points for $P_0(N,\tau)$ are obtained running a circuit 
equivalent to that of the Figure \ref{Circ_x}, with different values of $N$, on the simulator QISKIT of IBM on a classical 
computer and doing $90000$ simulations for each point. The continuous line for $P_0(N,\tau)$ is a plot  of equation (\ref{P_0T})
for $\theta=0$, i. e. 
$$P_0(N,\tau)= \frac{1}{2} \left(1+\cos^N(2\omega\tau)\right),$$
and is in good 
agreement with the {\it experimental} results of the quantum circuit for all values of $N$.

Actually it is easy to see that our circuit in figure 1 with $N$ ancillas, i. e. number of cycles 
$n=1$, and $\bra\alpha U\ket\alpha=0$ (a generalization of the previous situation $\ket\alpha=\ket0$ 
and $U=X$) is equivalent to a version of the Zeno dynamics in which after measuring the state 
in the computational basis it collapses to one of the eigenstates of the apparatus, $\ket0$ or 
$\ket1$, but in both cases it continues to evolve (i. e. in this case we do not project to $\ket 0$). The quantum circuit can be depicted:
\begin{figure}[h]
 \center
  \includegraphics[width=15cm]{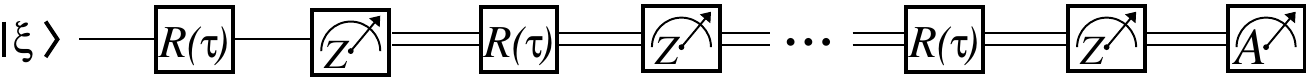}
\label{graf_ComparaZ_CX}
\end{figure}

The lesson that we extract from this equivalence is that we can replace multiple measurements, inherent to the Zeno effect, 
by the unitary evolution of the system coupled to multiple ancillas and a single measurement at the end. The important 
property of freezing the evolution is the same in both cases. Note in passing that
this equivalence implies the inequality 
$P_0(N,\tau)>Z_0(N,\tau)$, that we noticed before. In fact, in the Zeno process, $Z_0$, we drop out the cases in which after measuring in 
the computational basis we get $\ket1$ while in our case, $P_0$, we do not. It is then clear that our circuit has 
more probability of producing $\ket0$ at the end than for the Zeno case. 

In the rest of the section we shall extend our results for $\bra\alpha U\ket\alpha\not=0$. In order 
to keep the derivation relatively simple we will particularize to the evolution $R_\theta(\tau)$ with $\theta=0$. 
We simplify the notation by denoting $c=\cos(\omega\tau)$ and $s=\sin(\omega\tau)$. Therefore
 \begin{eqnarray}
  R_0(\tau)= \left( \begin{array}{c c}
             c&-s\\s&c
            \end{array}
\right).
 \end{eqnarray}
 We shall take the qubit $\ket{q_0}$ in the state $\ket0$ and all the ancillas in the same state 
 $\ket\alpha$. The conditioned gates are all the same $C_U(0,i)$ with arbitrary unitary operator $U$.
 
 Hence, we are in the situation of sec. II with $n=1$ and we can apply the recursion relation (\ref{recurrence})
 with 
\begin{equation}{\cal M}=\left(\begin{array}{c c c c}
             c^2 & s^2 & - c s & -c s\\
             s^2 & c^2 & c s & c s\\
             z c s & -z c s & z c^2 & -z s^2\\
             {z^*} c s & -{z^*} c s & -{z^*} s^2 & {z^*} c^2
           \end{array}
\right), 
\end{equation}
where $z=\bra\alpha U\ket\alpha$, as before.

We want to compute the probability $P_0(N,\tau)$ for the qubit $\ket{q_0}$  to be in the state 
$\ket{0}$  when a time $T=N\tau$ has elapsed. Given the initial data for the recurrence (\ref{recurrence}) 
$$\normsq{\phi_{0}}=1,\quad \normsq{\eta_{0}}=
 \sprod{\phi_{0}}{\eta_{0}}=
 \sprod{\eta_{0}}{\phi_{0}}=0,$$
it is clear that $P_0(N,\tau) = |({\cal M}^N)_{1,1}|^2$. As we are interested in the limit $\tau <<1$, the matrix element
$({\cal M}^N)_{1,1}$ can be calculated, up to 
second order in $\tau$ using the spectral decomposition of  $\cal M$ and this results in:
\begin{equation}
 P_0(N,\tau)= 1-\omega^2\tau^2 \left(\frac{N (1-z {z^*})}{(1-z)(1- z^*)}+
 \frac{z^{N+1}-z}{(1-z)^2}+\frac{{z^*}^{N+1}-{z^*}}{(1-{z^*})^2}\right)+O(\tau^4).
 \label{P0mul}
\end{equation}
As a first example we can consider the case $\ket\alpha=\ket0$ and
$U=X$ where we recover the previous situation for $\theta=0$. 
In fact, given that $X\ket0=\ket1$ and $z=\bra0 X\ket0=0$ we obtain
$$P_0(N,\tau)= 1-N\omega^2\tau^2+O(\tau^2)$$
as can be derived from the expansion of the exact result  (\ref{P_0T}).

A more general situation is when  $\bra\alpha U\ket\alpha=z$ is a real number, in this case (\ref{P0mul}) results in:
\begin{equation}
 P_0(N,\tau)= 1-\omega^2\tau^2 \left(\frac{N(z+1)}{1-z}+2 z\frac{z^N-1}{(1-z)^2}\right)+O(\tau^4),
\end{equation}
that for $N>>1$ and $z<1$ gives 
$$P_0(N,\tau)= 1-\omega^2\tau^2 \left(\frac{N(z+1)}{1-z}-\frac{2 z}{(1-z)^2}\right)+O(\tau^4).$$

\begin{figure}[h]
 \center
  \includegraphics[width=10cm]{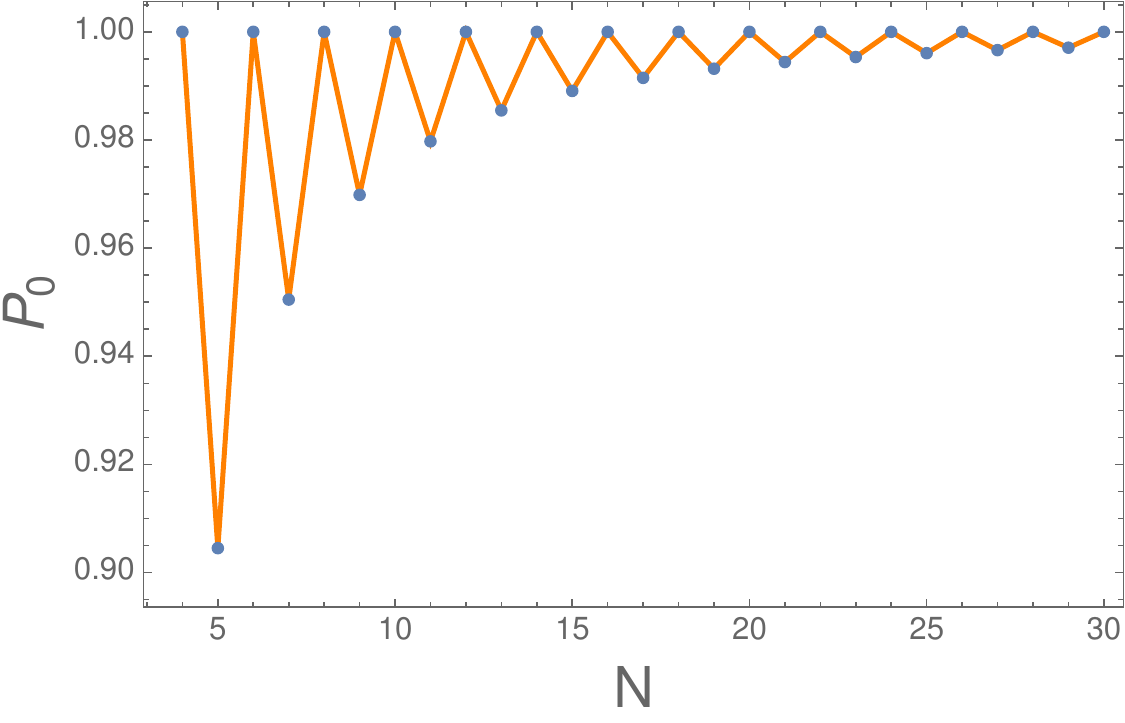}
  \caption{Probability $P_0(N,\tau)$ of finding 
  the qubit $\ket{q_0}$ in the state $\ket{0}$ as a function of the number of intervals $N$ that we divide the total 
   time $T$, for $\omega T=\pi/2$, and using $C_z$ as the controlled gate. All the ancillas are in the state $\ket{1}$ and 
   $\ket{\xi}=\ket{0}$.} 
\label{CZ_tot}
 \end{figure}

Another interesting case is when 
$\bra\alpha U\ket\alpha=z=\exp(\ii\varphi)$, in this case 
\begin{equation}
 P_0(N,\tau)= 1-\omega^2\tau^2 \frac{\sin^2(N \varphi/2)}{\sin^2(\varphi/2)}+O(\tau^4),
\end{equation}
and in particular,  when $\varphi=\pi$ we obtain that 
\begin{eqnarray}
 P_0(2 N,\tau)&=&1,\\
P_0(2 N+1,\tau)&=&1-\omega^2\tau^2+O(\tau^4),
\end{eqnarray}
which corresponds to take $U=Z$ and the initial state of the ancillas
$\ket\alpha=\ket{1}$. 
In this case it is easy to solve exactly the recursion relations (\ref{itera1}) 
since for all $k$ we have that
\begin{eqnarray}
 \ket{\Psi_{2 k}}&=& \ket{0}\otimes\ket{1}^N =\ket{\Psi_0}  ,\label{sz1}\\
 \ket{\Psi_{2 k+1}}&=& (\cos(\omega\tau) \ket{0}-\sin(\omega\tau))\ket{1})\otimes\ket{1}^N  , \label{sz2}
\end{eqnarray}
this result means that after any  even number of steps the system returns to the initial state and $P_0(2 k,\tau)=1$ 
independently of the integer value of $k$. In these circumstances, after any odd number of steps and taking $\omega T=\pi/2$, we have
$P_0(2k+1,\tau)=\cos^2(\omega\tau)=\cos^2(\pi/(2 N))$. Figure \ref{CZ_tot} represents the probability $P_0(N,\tau)$ when 
the controlled gate is $C_Z(0,i)$, for $\omega T=\pi/2$ and the time $T$ is divided in $N$ intervals.  We can see 
that $P_0(N,\tau) > 0.9$ for any $N>3$ and also the fact that $P_0(2 k,\tau) =1$ for any integer $k$.
 Note that because the recursion relations depend only of the initial conditions and 
  the value of the scalar product $\sprod{\alpha}{U |\alpha}$, these 
 results obtained  putting all the ancillas in the the state 
 $\ket{1}$  and using $C_{Z} (0,i)$ as 
 the controlled gate, are the same that we would have obtained if all the ancillas where in the state 
 $\ket{-}=\frac{1}{\sqrt{2}} (\ket{0}-\ket{1})$ and we would have used the controlled gate $C_X(0,i)$.
 It is important to remark that in the $C_Z$ case, the state $\ket{1}$ 
 of the ancilla that serves as the target for the 
 the controlled $C_{Z}$ operation does not change and that the final 
 state of all the ancillas is the same as the 
 initial state $\ket{1}^N$. This means that the neat effect of the 
 $N$ ancillas can be substituted by just only 
 one ancilla. The situation will be the same whenever the initial 
 state $\ket{\alpha}$ is an eigenstate 
 of the unitary operator used in the controlled gate, 
 with an eigenvalue different from $1$, 
 ($U \ket{\alpha} = \ee^{i \varphi}\ket{\alpha}, \; \varphi \ne 0$). In that situation, 
 if we use only one ancilla, after each 
 controlled operation the ancilla will remain in the same eigenstate and can be used for 
 the next operation in the same conditions as before. This case is studied in the following section.
 \section {Simulation with one auxiliary qubit}
\begin{figure}[h]
 \center
  \includegraphics[width=10cm]{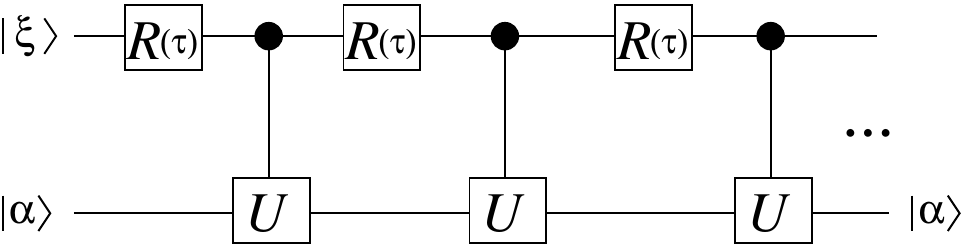}
  \caption{Circuit for a simulation using only one ancilla. $R(\tau)$  represents a rotation on the space of states of
  $\ket{q_0}$ and $U$ is a general unitary gate. The state $\ket{\alpha}$ is an eigenvector of $U$ with eigenvalue 
  $e^{i\varphi}$ and $\varphi\ne 0$.}
\label{Zenon_U1}
 \end{figure}

 The results of the previous section show that we can simulate the QZE by coupling the qubit $\ket{q_0}$ to 
 $N$ auxiliary ancillas which, in some sense, make the effect of the coupling of $\ket{q_0}$ to a macroscopic system. 
 However, in terms of controlling a qubit  this method is unpractical when $N\gg 1$ since 
 we would need a very high number of ancillas if we want to control only a few qubits. 
 Then it is better to use the results obtained in the last part of the previous  section 
 to simulate the QZE using only one auxiliary ancilla, so 
 now we consider a circuit with the qubit of interest $\ket{q_0}$ and only one ancilla $\ket{q_1}$. 
 We shall alternate the rotations $R_0(\tau)$ of $\ket{q_0}$ with a controlled unitary operation $C_U(0,1)$ as shown in 
 Figure \ref{Zenon_U1}. The initial states are $\ket{q_0}=\ket{0}$ and $\ket{q_1}=\ket{\alpha}$ where $U \ket{\alpha}=
 \ee^{i\varphi} \ket{\alpha}$ and $\varphi\ne 0$, (note that if $\varphi =0$
 the gate $C_U (0,i)$ will act as the identity).
 The recursion relations for the system can be obtained as before.  Suppose that after $k$ steps the wave function of the system is 
 \begin{equation}
   \ket{\Psi_k}= 
   (a_k \ket{0} + b_k \ket{1})\ket{\alpha},
 \end{equation}
then after the next step 
\begin{eqnarray}
  \ket{\Psi_{k+1}} &=& C_U(0,1) R_0(\tau) \ket{\Psi_k}\cr &=& 
  \left( (\cos(\omega\tau)\, a_k -\sin(\omega\tau)\, b_k) \ket{0} + 
  \ee^{\ii \varphi} (\sin(\omega\tau)\, a_k  + \cos(\omega\tau)\, b_k )\ket{1} \right)\ket{\alpha},
  \end{eqnarray}
or equivalently
\begin{equation}
 \left(\begin{array}{c}
        a_{k+1}\\b_{k+1}
       \end{array}
\right) = T^k(\tau, \varphi) 
\left(\begin{array}{c}
        a_{1}\\b_{1}
       \end{array}
\right),\;\;\; 
T(\tau,\varphi)=
\left(\begin{array}{c c}
        \cos(\omega\tau)& -\sin(\omega\tau)\\
        \ee^{\ii \varphi}\sin(\omega\tau)&\ee^{\ii \varphi}\cos(\omega\tau)
       \end{array}
\right), 
\label{recurrence1a}
\end{equation}
with the initial conditions $a_0=1,\; b_0=0$. As we are interested in the behavior of 
$P_0(N,\tau),P_1(N,\tau)$ for $\tau << 1$ (frequent interactions)
we can expand $T^N(\tau,\varphi)$ up to second order in $\tau$ and find:
\begin{eqnarray}
  &&T^N_{1,1}(\tau,\varphi)=\left(1-\ii\frac{\omega^2\tau^2}2\cot(\varphi/2)\right)^N\!-\omega^2\tau^2\;\frac{1-\ee^{\ii N \varphi}} {4 \sin^2(\varphi/2)} + O(\tau^3)\\[0.8mm]
  &&T^N_{1,2}(\tau,\varphi)= \omega\tau \;\frac{1-\ee^{\ii N \varphi}}{\ee^{\ii  \varphi}-1}- \ii N\omega^3\tau^3\cot(\varphi/2)
  \frac{1+\ee^{\ii N \varphi}}{\ee^{\ii  \varphi}-1}+O(\tau^3),\\[0.8mm]
  &&T^N_{2,1}(\tau,\varphi)= \omega\tau \;\frac{1-\ee^{\ii N \varphi}}{\ee^{-\ii  \varphi}-1}- \ii N\omega^3\tau^3\cot(\varphi/2)
  \frac{1+\ee^{\ii N \varphi}}{\ee^{-\ii  \varphi}-1}+O(\tau^3),\\[0.8mm]
  &&T^N_{2,2}(\tau,\varphi)=\ee^{\ii N \varphi}\left(1+\ii\frac{\omega^2\tau^2}2\cot(\varphi/2)\right)^N\!+\omega^2\tau^2\;\frac{1-\ee^{\ii N \varphi}} {4 \sin^2(\varphi/2)} + O(\tau^3).
 \end{eqnarray}
The probability of finding $\ket{q_0}$ in the state $\ket{0}$ after 
$N$ steps, when the initial state is $\ket{0}$ will be 
\begin{equation}
 P_0(N,\tau)=|a_N|^2 = |T^N_{1,1}(\tau,\varphi)|^2 = 1-\omega^2\tau^2 \,\frac{\sin^2(N \varphi/2)}{\sin^2(\varphi/2)} +O(\tau^4).
 \label{P0series}
\end{equation}
\begin{figure}[t]
 \center
  \includegraphics[width=10cm]{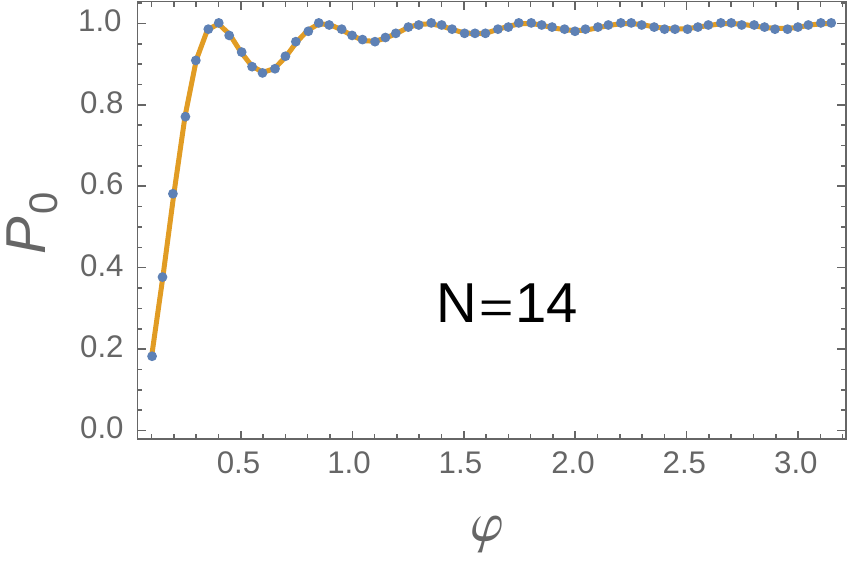}
  \caption{Exact results for $P_0(N=14,\tau)$ as a function of $\varphi$, presenting  peaks  when the angle 
  is a multiple of $\frac{2 \pi}{N}$. Here  $\omega T= \pi/2$,  $\tau=T/N$ and $\ket{\xi}=\ket{0}$.}
\label{fase1}
 \end{figure}
The equation (\ref{P0series}) is an expansion of $P_0(N,\tau)$ in terms of $\tau=(\frac{T}{N})$, but it should be used 
carefully for small values of $\varphi$. In fact for $\varphi \to 0$ the coefficient of the $\tau^{n}$ 
term grows as $N^{n}$ and the expansion is not valid any more. This corresponds to the fact that for $\varphi=0$ the controlled unitary operation 
acts as the identity and the probability $P_0(N,T)=\cos^2(\omega T)$ independent of $N$.

For $\varphi\not=0$, the equation
(\ref{P0series}) shows that $P_0(N,\tau)$ has maximums when $N \varphi $ is a multiple of $2 \pi$, this is a generalisation 
of the behavior of $P_0$ that we found in the previous section  using the $C_Z$ gates.
In the Figure \ref{fase1} we represent the exact results, obtained numerically by solving the recurrence 
relations (\ref{recurrence1a}) for the case $N=14$ and different values of $\varphi$. There one can see the maxima of $P_0(14,\tau)$ 
correspond to values of $\varphi$ multiple of $\frac{2 \pi}{N}$. It is also 
interesting to remark that due to the quadratic form of the corrections, for values of 
$\varphi$ of order 1, the limiting value is very soon approached. For instance,  $P_0(14,\tau) \ge 0.95$ for all values of $\varphi\ge 1.0 $. 
\begin{figure}[h]
 \center
  \includegraphics[width=10cm]{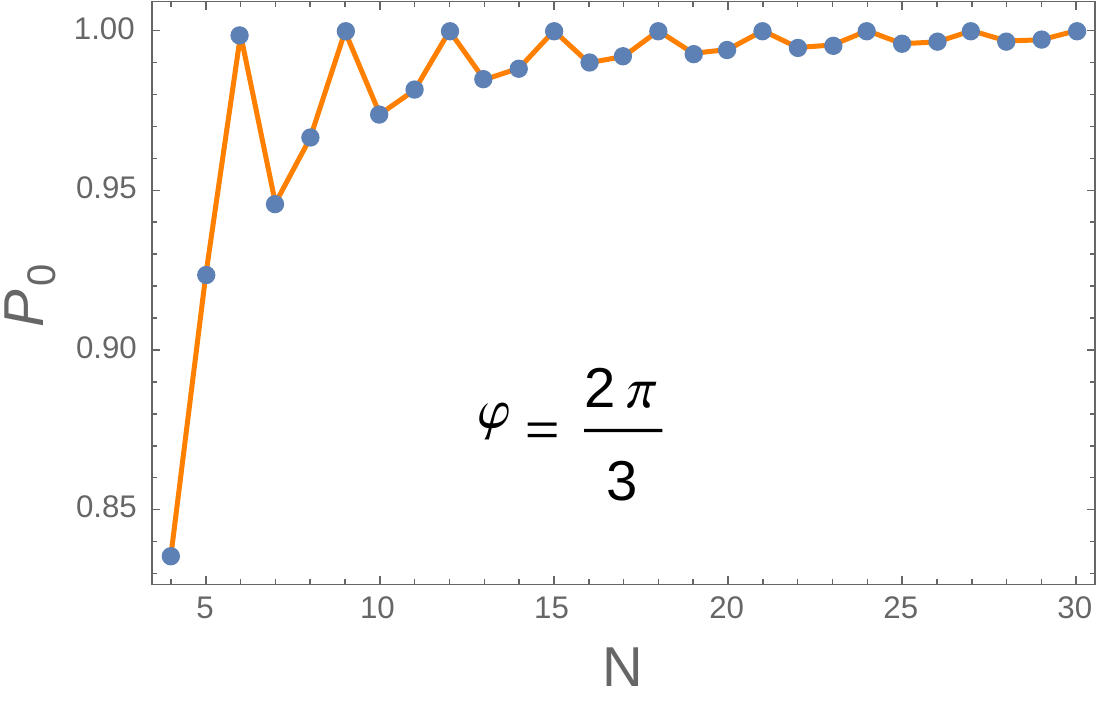}
  \caption{Exact results for $P_0(N,\tau)$ for  $\varphi=\frac{2 \pi}{3}$ and different values of $N$, presenting  maxima when  
  $N$ is a multiple of $3$. As before, $\omega T=\pi/2$, $\tau=T/N$ and $\ket{\xi}=\ket{0}$.}
\label{fase2}
\end{figure}

The points in Figure \ref{fase2} 
are the exact results for $P_0(N,\tau)$ as a function of $N$ when $\varphi=2 \pi/3$ and, as expected,  
it has maxima for values of  $N$ multiple of $3$. This, again, generalizes the results for the $C_Z$ 
gate where the peaks of  $P_0(N,\tau)$  happen for all even values of $N$.

\section{Conclusions}\label{sec_conclu}
We have studied the simulation of the Zeno effect on a quantum computer. 
To do so, we use the property that the repeated action
of a unitary operator on a system can simulate the  coupling of this system with another macroscopic one, then inducing 
an evolution analogous to that of the Zeno effect. We have performed the analysis using two different strategies. In the first one, 
the macroscopic system is represented by a big number of ancillas, all of them coupled to our system of interest $\ket{q_0}$ 
by means of unitary operations controlled by the qubit $\ket{q_0}$ and acting, each time, on a different ancilla. In this case 
we have obtained the exact evolution of $\ket{q_0}$ when the controlled operations were of the type $C_X(0,i)$ or $C_Z(0,i)$. 
For more general unitary operations we have derived the final state up to second order in the parameter $\tau=T/N$. 
In all cases the continued action of the unitary operations freezes the evolution of $\ket{q_0}$ in a way that mimics the Zeno effect.

These results can be used for the stabilization of the physical qubits of a quantum computer. In this case it is more economical to use a single ancilla to which the controlled gate repeatedly couples. This has been implemented taking the ancilla in an eigenstate $\ket{\varphi}$ of the unitary operator used in the controlled operations. This method results very efficient, in particular 
for certain values of $\varphi$, and in general produces a good simulation of the Zeno effect.

\begin{acknowledgments}
We thank Paolo Facchi and Kazuya Yuasa for discussions and for bringing some useful references to our attention.
We are partially supported
by Spanish Grants No. PGC2022-126078NB-C21 funded
by MCIN/AEI/10.13039/
501100011033,
the Quantum Spain project of the QUANTUM ENIA of the
 Ministry of Economic Affairs and Digital Transformation, the Diputaci\'on General de
Arag\'on-Fondo Social Europeo (DGA-FSE) Grant No. 2020-E21-17R of the Arag\'on Government, and the European Union, NextGenerationEU Recovery and
Resilience Program on  ``Astrof\'{\i}sica y F\'{\i}sica de Altas Energ\'{\i}as", CEFCA-CAPA-ITAINNOVA.
\end{acknowledgments}


\begin{thebibliography}{XXX}
    
\bibitem{Sudarshan} B. Misra and E. C. G: Sudarshan
\textit{The Zeno's paradox in quantum theory}, 
Jour. of Math. Phys. 18, 756 (1977); 
see also and 
H. Ekstein and A. Seigert,
{\it On a reinterpretation of decay experiments}, Ann. Phys. (N.Y.) 68, 509-520 (1971)
\bibitem{Pascazio1} P. Facchi and Saverio Pascazio
\textit{Quantum Zeno dynamics: mathematical and physical aspects},
 arXiv:0903.3297v1 [math-ph] , and J. Phys. A: Math. Theor. 41 (2008) 493001.
\bibitem{Pascazio} H. Nakazato, M. Namiki, and S. Pascazio
\textit{Temporal behavior of quantum mechanical systems}, Int. J. Mod. Phys. B 10, 247 (1997)
; Saverio Pascazio
\textit{All you ever wanted to know about the quantum Zeno effect
in 70 minutes},
arXiv:1311.6645v1, and Open Sys. Inf. Dyn.
21, 1440007 (2014).
\bibitem{Whitaker} D. Home and M. A. Whitaker, 
\textit{A conceptual analysis of quantum Zeno; paradox, measurement, and experiment}, 
Ann.  Phys (NY) 258,237 (1997).
\bibitem{Shimizu}
K. Koshino and A. Shimizu
\textit{ Quantum Zeno effect by general measurements},
Phys. Rep. 412, 191 (2005).
\bibitem{flp}P. Facchi, D. A. Lidar and S. Pascazio,
\textit{Unification of dynamical decoupling and the quantum Zeno effect},
Phys. Rev. A {\bf 69}, 032314, (2004).
\bibitem{bfgk} Daniel Burgarth, Paolo Facchi, Giovanni Gramegna and Kazuya Yuasa,
\textit{One bound to rule them all: from Adiabatic to Zeno},
Quantum 6, 737 (2022), and arXiv, quant-ph: 2111.08961v2 (2022)
\bibitem{Haroche}
J. D. Franson, B. C. Jacobs and T. B. Pittman 
\textit{Quantum computing using single photons and the Zeno effect}, 
Phys. Rev. A {\bf 70} 062302 (2004);
Gerardo A Paz-Silva, A. T. Rezakhani, Jason M. Dominy and D. A. Lidar 
\textit{Zeno effect for quantum computation and control},
Phys. Rev. Lett. 108, 080501 (2012); 
J. M. Raimond, P. Facchi, B. Peaudecerf, S. Pascazio, C. Sayrin, I. Dotsenko, S. Gleyzes, 
M. Brune and S. Haroche, 
\textit{Quantum Zeno dynamics of a field in a cavity},
Phys. Rev A {\bf 86} 032120 (2012)..
\bibitem{Panigrahi} Subhashish Barik, Dhiman Kumar Kalita, Bikash K. Behera and Prasanta K. Panigrahi
\textit{Demonstrating Quantum Zeno Effect on IBM Quantum Experience},
arXiv:2008.01070v1
 \end{thebibliography}
 \end{document}